# The IQIYI System for Voice Conversion Challenge 2020

*Wendong Gan*[1], *Haitao Chen*[1], *Yin Yan*[1], *Jianwei Li*[1] ,

*Bolong Wen*[1], *Xueping Xu*[1], *Hai Li*[1]

[1]IQIYI Inc.

ganwendong@qiyi.com

## Abstract

This paper presents the IQIYI voice conversion system (T24) for Voice Conversion 2020. In the competition, each target speaker has 70 sentences. We have built an end-to-end voice conversion system based on PPG. First, the ASR acoustic model calculates the BN feature, which represents the content-related information in the speech. Then the Mel feature is calculated through an improved prosody tacotron model. Finally, the Mel spectrum is converted to wav through an improved LPCNet. The evaluation results show that this system can achieve better voice conversion effects. In the case of using 16k rather than 24k sampling rate audio, the conversion result is relatively good in naturalness and similarity. Among them, our best results are in the similarity evaluation of the Task 2, the 2nd in the ASV-based objective evaluation and the 5th in the subjective evaluation.

## 1. Introduction

In order to improve the technology of voice conversion, the voice conversion challenge has been held every two years since 2016 [1][2]. The competition is trained on the same data set. Voice conversion technology has also developed rapidly in recent years. Initially, people studied speech conversion technology based on parallel corpus, and achieved certain results. Such as Sprocket [3], Merlin [4]. However, because it is convenient to obtain parallel corpus in actual use, this method is difficult to be applied in practice. Therefore, people began to study non-parallel corpus. At present, non-parallel corpus speech conversion systems are mainly divided into: PPG-based speech conversion systems, phoneme decomposition-based speech conversion systems [5], and GAN-based speech conversion systems [6][7]. Each system has its own characteristics. In the case of actual commercial use, PPG-based voice conversion systems are more selected and developed. Its PPG feature is obtained by ASR's acoustic model. This module has been developed more maturely in the industry, and thousands or even tens of thousands of hours of data have been seen during the training process, and data has been enhanced. Therefore, the versatility of this ASR data can also make the trained voice conversion system have better practical versatility. It can also resist interference better.

For cross-language conversion, for example, the target speaker has only Chinese corpus, and he needs to convert the source voice of English. This is a new branch of voice conversion developed in recent years. Due to the development of internationalization, people's demand for this kind of cross-language conversion is also increasing. This is also entertaining. A more common solution is to recognize the English source speech through ASR to obtain the text, and then perform certain processing on the text, and use the model trained on the target corpus to obtain the voice of the target speaker. This approach will cause more pronunciation accuracy problems, there may be more inconsistencies in the input and output audio content, and there may also be greater inconsistencies in timing. The effect will be very limited. In order to improve, researchers have proposed an end-to-end cross-language conversion scheme based on PPG features [8]. At present, in order to improve the effect, everyone mainly includes the following three methods: using dual language phoneme tables, splicing the PPG results of dual voice ASR, and using PPG's Bottleneck feature. In contrast, the last method can obtain relatively good results, and the demand for the ASR acoustic model is relatively lower. It is expected that this will be the most widely adopted plan at this stage.

Many aspects of the voice conversion system come from speech synthesis. With reference to speech synthesis, since it is still difficult to directly model time series audio, the more common mode at this stage is still the acoustic model and vocoder. The LPCNet [9][10] vocoder has better real-time performance under the condition that a higher MOS score can be obtained, and has been researched and improved by many scholars. It is based on LPC features, but Mel and Linear features are more commonly used in the industry, so LPC features are more troublesome to maintain and use. In this system, we successfully optimized the LPCNET vocoder into a Mel-based vocoder, namely Mel- LPCNet.

Following the latest development of the voice conversion system, we proposed the IQIYI system, which can complete Task 1 and Task 2 of VCC2020 at the same time. In the training phase, first, we calculate the Bottleneck feature of PPG through the ASR acoustic model; obtain the Mel feature through signal processing; train the conversion model of BN and Mel. In the inference stage, we extract BN features from the source speech, and obtain Mel features through the trained conversion model. In order to improve the effect, we added prosody embedding to increase the stability of the system and the similarity of the synthesized sound. Experiments show that the above improvements have achieved relatively good results. Finally, we train the Mel-LPCNet vocoder. In the conversion phase, we convert Mel to wav to complete the voice conversion.

The organization of this article is as follows: Part 2: Introduce the tasks of VCC2020 in detail; Part 3: Demonstrate the specific improvement implementation methods of our system; Part 4: Introduce the system's evaluation results and experimental comparison. Finally, the conclusion and future work will be described in the fifth section.

## 2. The tasks in Voice Conversion 2020

There are two tasks this year. Compared with the previous two sessions, Task 2 cross-language conversion has been added. The two tasks are:

1) Task 1: voice conversion within the same language

In training, a sentence set uttered by the source speaker is different from that uttered by the target speaker, but they are still the same language. Moreover, only a small number of sentences are shared between these two sentence sets.

In conversion, the source speaker's voice is converted as if it is uttered by the target speaker while keeping linguistic contents unchanged.

We will provide voices of 4 source and 4 target speakers (consisting of both female and male speakers) from fixed corpora as training data. Each speaker utters a sentence set consisting of 70 sentences in English. Only 20 sentences are parallel and the other 50 sentences are nonparallel between the source and target speakers.

Using these data sets, voice conversion systems for all speaker-pair combinations (16 speaker-pairs in total) will be developed by each participant.

2) Task 2: cross-lingual voice conversion

In training, a sentence set uttered by the source speaker is totally different from that uttered by the target speaker as a language of the source speaker is different from that of the target speaker.

In conversion, the source speaker's voice in the source language is converted as if it is uttered by the target speaker while We will also provide voices of other 6 target speakers (consisting of both female and male speakers) from fixed corpora as training data. The source speakers are the same as in the 1st task. Each target speaker utters another sentence set consisting of around 70 sentences in a different language; 2 target speakers utter in Finnish, 2 target speakers utter in German, and 2 target speakers utter in Mandarin.

Using these nonparallel data sets, voice conversion systems for all speaker-pair combinations (24 speaker-pairs in total) will be developed by each participant ping linguistic contents unchanged.

## 3. IQIYI Voice Conversion System

### 3.1、Framework

As indicated in Figure 1, our IQIYI Voice Conversion System consists of two parts, the training phase and the conversion phase.

In the training phase, we get the Mel spectrogram through SFFT preprocessing. We get Bottleneck features through the ASR acoustic model, which contains things related to speech content and retains time series information. In order to solve the problem of cross-language voice conversion, we chose Bottleneck feature. The paper [8] has demonstrated that Bottleneck features has better cross-language capabilities than PPG. This has also been demonstrated in our Task 2 evaluation. Get Prosody embedding through the prosody reference encoder, which contains information related to rhythm, emotion, and speaker ID. After training, get the voice conversion model.

In the inference voice change stage, Mel Spectrogram, Bottleneck, and prosody Embedding features are obtained respectively. And then, the converted Mel Spectrogram is obtained through the Voice Conversion model. Finally, the trained Mel-LPCNet vocoder generates audio wav of voice conversion.

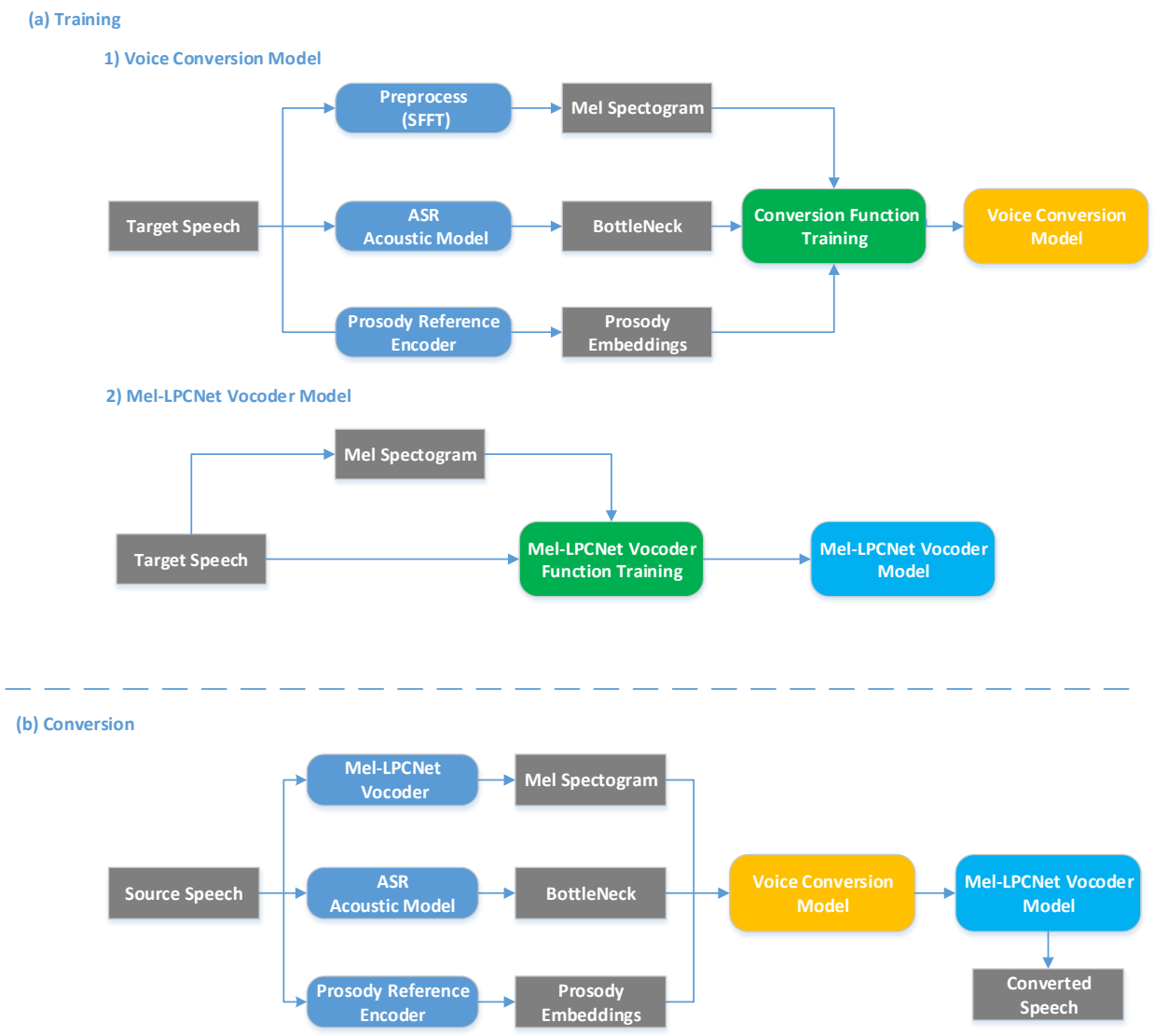

Figure 1: The framework of the proposed system.

### 3.2、Data processing

We used the LJSpeech dataset as the training data for the base model, which is a single speaker reading passages and has a total length of approximately 24 hours. Downsample the 22.05k data to 16k. For the target speaker of the competition, each person has 70 sentences, about 3mins. We changed the data from a downsample of 24k to 16k. Considering that the amount of data for each person is very limited, we have done data enhancement. We change the speed without change pitch. The speed is 0.7, 0.8, 0.9 1.0, 1.1, 1.2, 1.3. That is, every target speaker actually has 420 sentences.

### 3.3、Voice Conversion Model

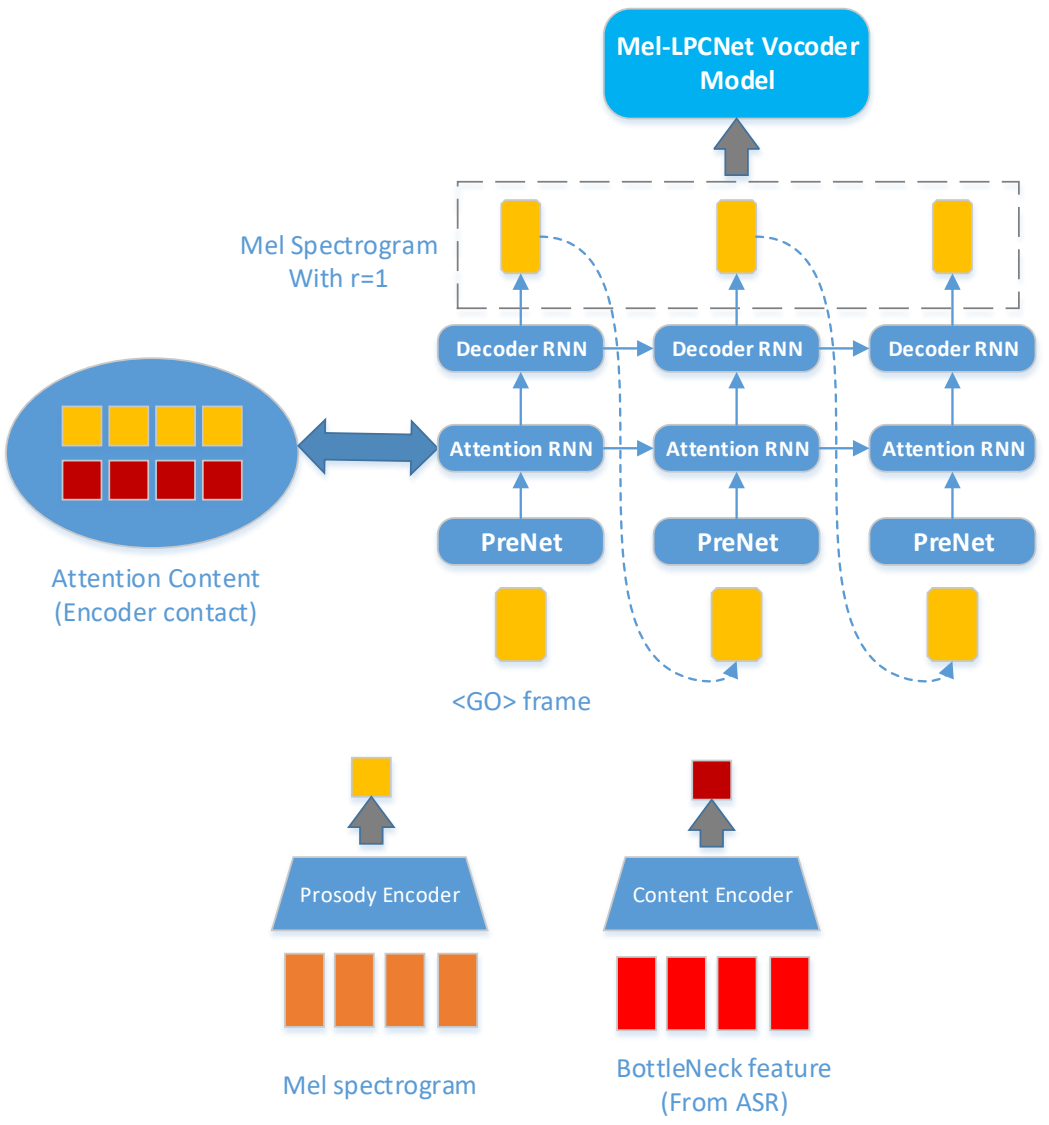

Figure 2: *The framework of the voice conversion model.*

As indicated in Fig 2, our system is mainly optimized based on Tacotron [11]. The Voice Conversion Model mainly contains three modules: Encoder, Decoder, and Attention. The specific parameters are shown in Table 1. The optimization we made is to add the Prosody encoder [12], which is to increase the similarity of the voice conversion and the stability of the system. Experiments show that without the prosody module, the alignment of the model is difficult. The effect of this has also been withdrawn in the evaluation results of Part 4 later. We will analyze in Part 4. The specific structure of the Prosody encoder will be described in Section 3.4

Table 1: *Voice Conversion Model Parameters*

| Voice Conversion Model | |
|---|---|
| Mel spectral analysis | Pre-emphasis: 0.97, frame length:50 ms; frame shift: 10ms; window type: hann |
| Encoder CBHG | Conv1D bank: K=16, conv-k-128-ReLU<br>Max pooling: stride=1, width=2<br>Conv1D projection: conv-3-128-ReLU -> conv-3-128-ReLU<br>Highway net: 4 layers of FC-128-ReLU<br>Bidirectional GRU: 128 cells |
| Encoder pre-net | FC-256-ReLU->Dropout(0.5)-> FC-256-ReLU->Dropout(0.5) |
| Decoder pre-net | FC-256-ReLU->Dropout(0.5)-> FC-256-ReLU->Dropout(0.5) |
| Decoder RNN | 2-layer residual GRU (256 cells) |
| Attention RNN | 1-layer GRU (256 cells) |
| Reduction factor (r) | 1 |

3.4、Prosody Encoder Module

In order to increase the stability and similarity of the conversion system, we have added this module. This module currently has two modes. One is a fixed length [12] [13], and the other is a variable length [14]. Prosody with varying length can better fine-grained modeling of prosody, but it will also reduce the instability or similarity of the generated audio due to the richness and uncertainty of the audio during inference. Considering our mission, audio is not expressive So, we choose a fixed-length prosody. This is mainly used to model the speech style of the speaker, sentence level and above, not phoneme level or smaller granularity.

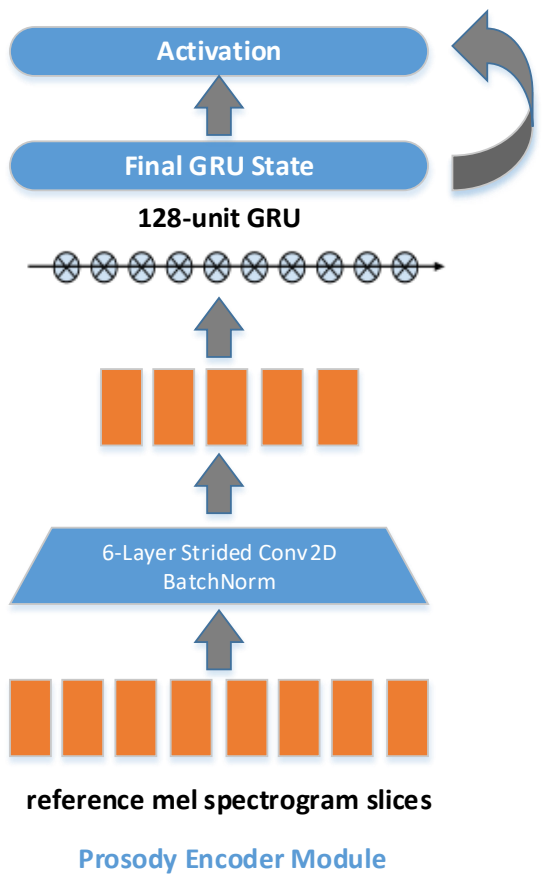

Figure 3: *The framework of the prosody encoder module.*

During training, we calculate a prosody encoder from the input of each sentence. When inferring, we calculate the prosody encoder of all training corpus, and then we select the cluster center sentence audio of each target speaker, and choose it as prosody embedding .

3.5、Mel-LPCNet Vocoder

For the vocoder, for the considerations of practical application, we chose LPCNet [9] instead of WaveNet [15]. This vocoder has good real-time performance when it can achieve better Naturalness MOS. In contrast, WaveNet's [15] real-time performance is too poor. At present, most popular vocoders use Mel spectrogram, while LPCNet vocoders use Bark-scale cepstral coefficients.. In order to have better practicability, we have improved the characteristics of LPCNet, using Mel, that is, we get Mel-LPCNet. Its structure is shown in the Figure 4. Unlike LPCNet, we calculate LPC based on the Mel spectrogram.

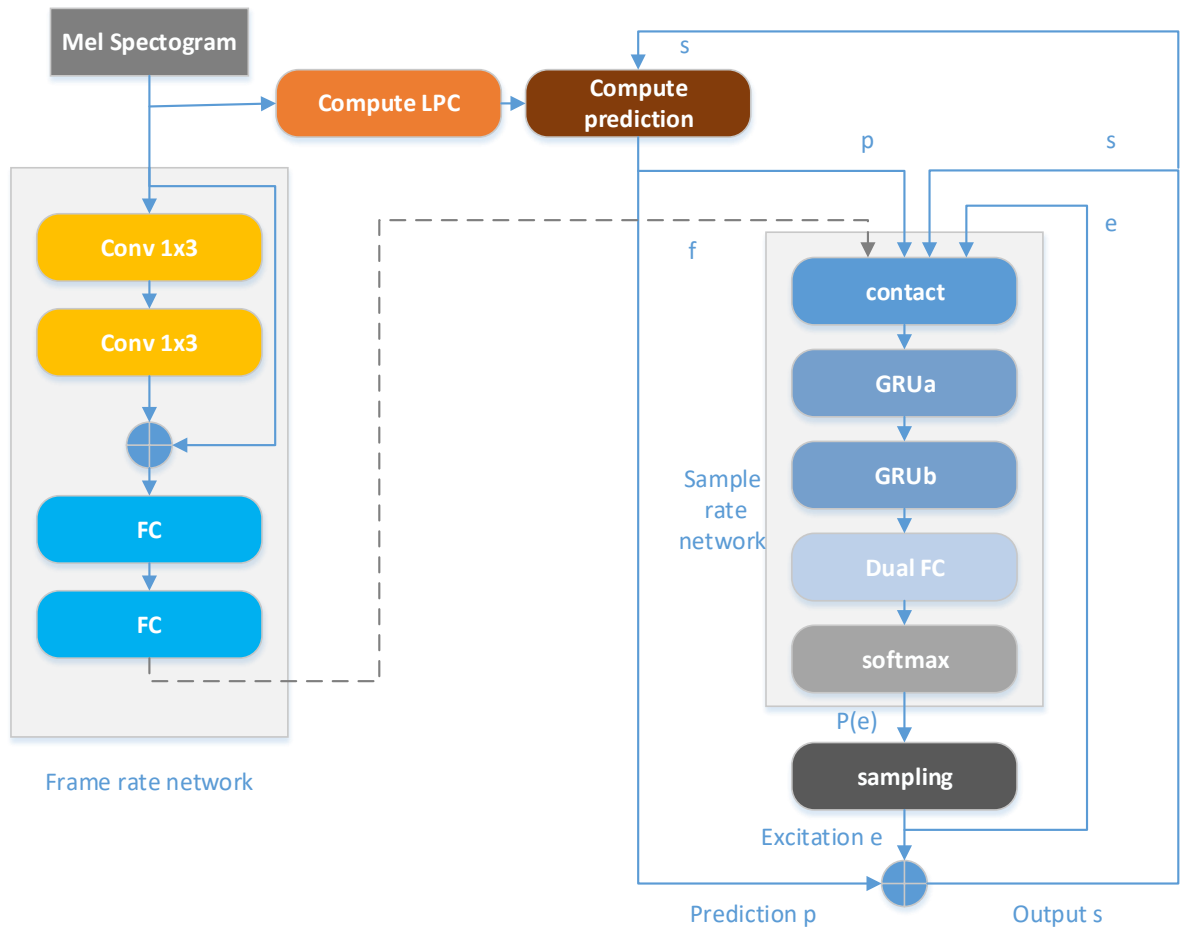

Figure 4: *The framework of the Mel-LPCNet vocoder.*

# 4. Evalution Results

Task 1 has 31 teams, including 3 baseline systems, one target audio, and one source audio. Task 2 has 28 participating teams, including 3 baseline systems, one target audio, and one source audio[16][17]. SOU is the source audio, which is the natural voice recorded. TAR is the target audio and also the recorded natural voice. T11, T16, T22 are baseline systems. T11 is Winner of Voice conversion challenge 2018. T16 is based on CycleVAE and Parallel WaveGAN. T22 is based on the ASR+TTS system. The other systems T1 to T33 are participating teams. System T24 is ours.

The official evaluation consists of two parts: Naturalness MOS (Mean opinion) and Similarity MOS. The Naturalness and Similarity MOS results are based on all the listeners' responses, including English Listeners and Japanese Listeners. In the case of using 16k rather than 24k sampling rate audio, the conversion result is relatively good in naturalness and similarity. Among them, our best results are in the similarity evaluation of the Task 2, the 2nd in the ASV-based objective evaluation and the 5th in the subjective evaluation. The performance of our system is as described later. Details are as follows.

## 4.1. Naturalness test

Figure 4 shows the results of the naturalness MOS given by all listeners for all the systems. In this test, listeners were asked to listen to samples and assign scores either on a scale of from 1 to 5: (1) Bad, (2) Poor, (3) Fair, (4) Good, and (5) Excellent. Our system has an average score of 3.9. We believe that if we increase the audio sample rate from 16 kHz to 24 kHz, our system can achieve a higher naturalness score.

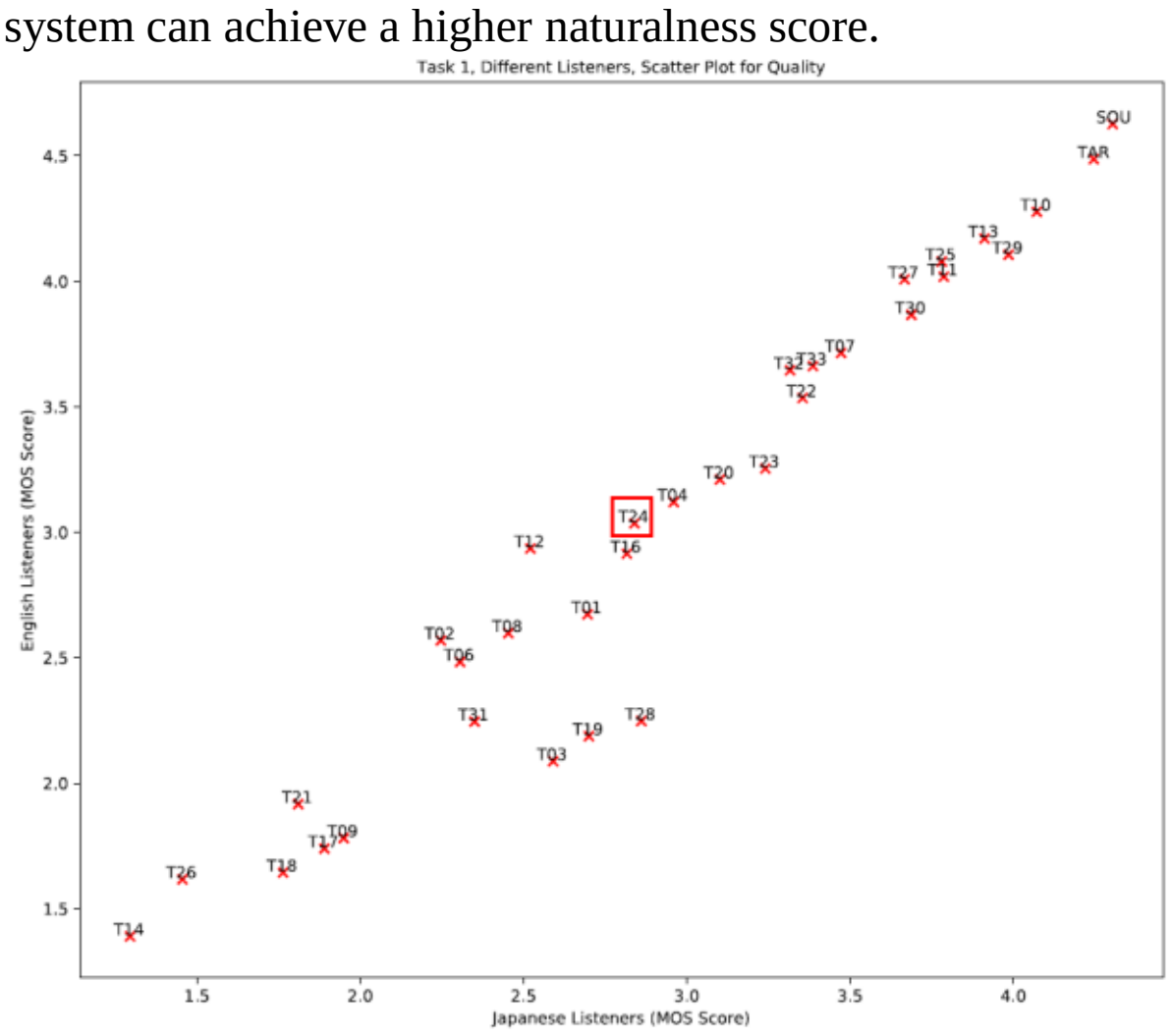

Figure 4: *Naturalness MOS evaluation of task 1, from both Japanese and English.*

### 4.2. Similarity test

Figure 6 and Figure 7 show the results of the similarity MOS given by both Japanese and English listeners for all the systems. In this test, listeners were asked to rate the speaker similarity of the two samples on a four-point scale: (4) same speaker, absolutely sure, (3) same speaker, not sure, (2) different speaker, not sure, (1) different speaker, absolutely sure. Again, the reference speech could be in English or either German, Finnish, or Mandarin for Task 2. The best results our system acquires is about 3.2 and ranks 5th. Compared with naturalness MOS, similarity MOS has achieved better results. This is mainly due to the powerful modeling capabilities of the prosody encoder module. Of course, We believe that if we increase the audio sample rate from 16 kHz to 24 kHz, our system can achieve a higher similarity score.

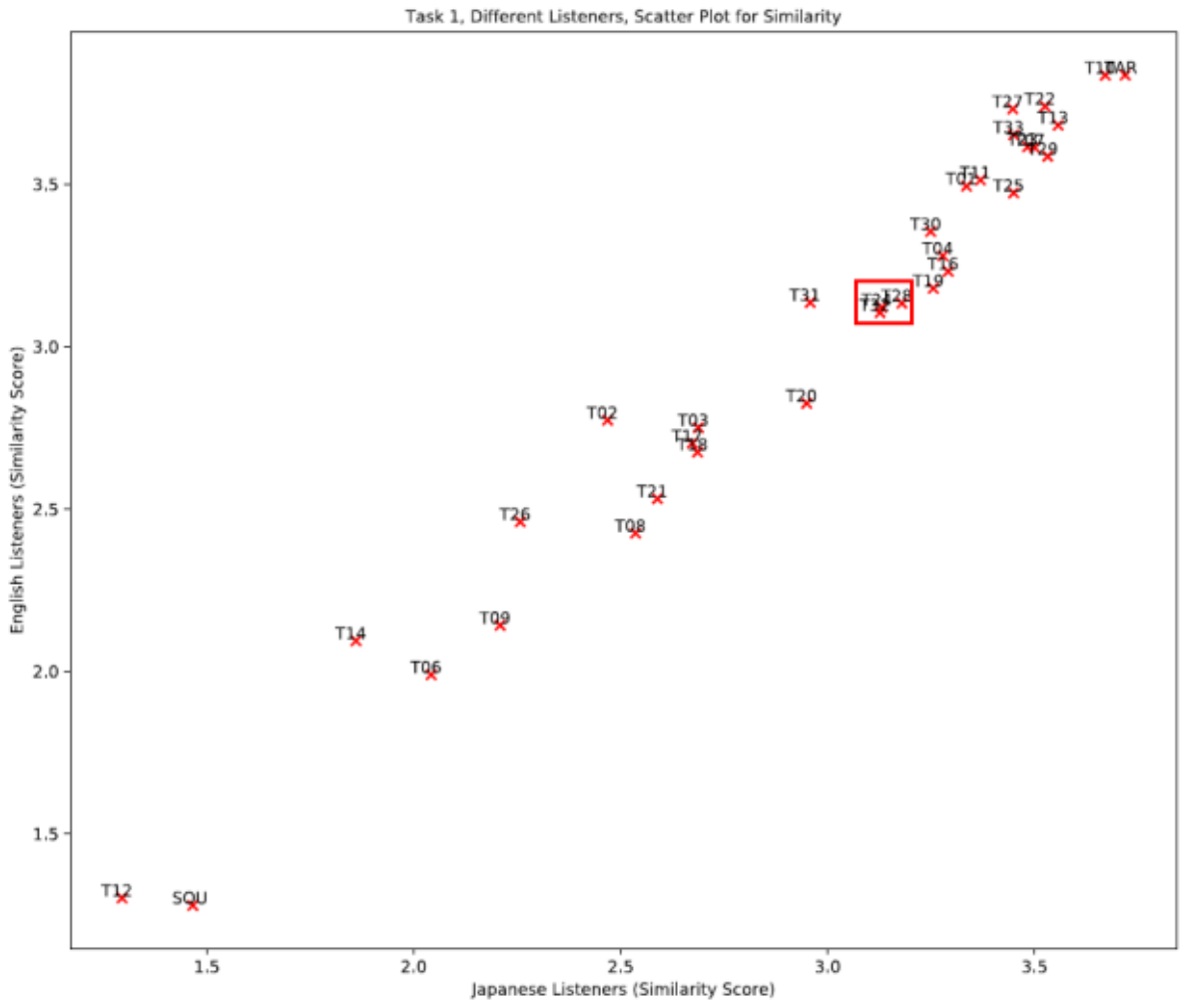

Figure 6: *Similarity MOS evaluation of task 1, from both Japanese and English.*

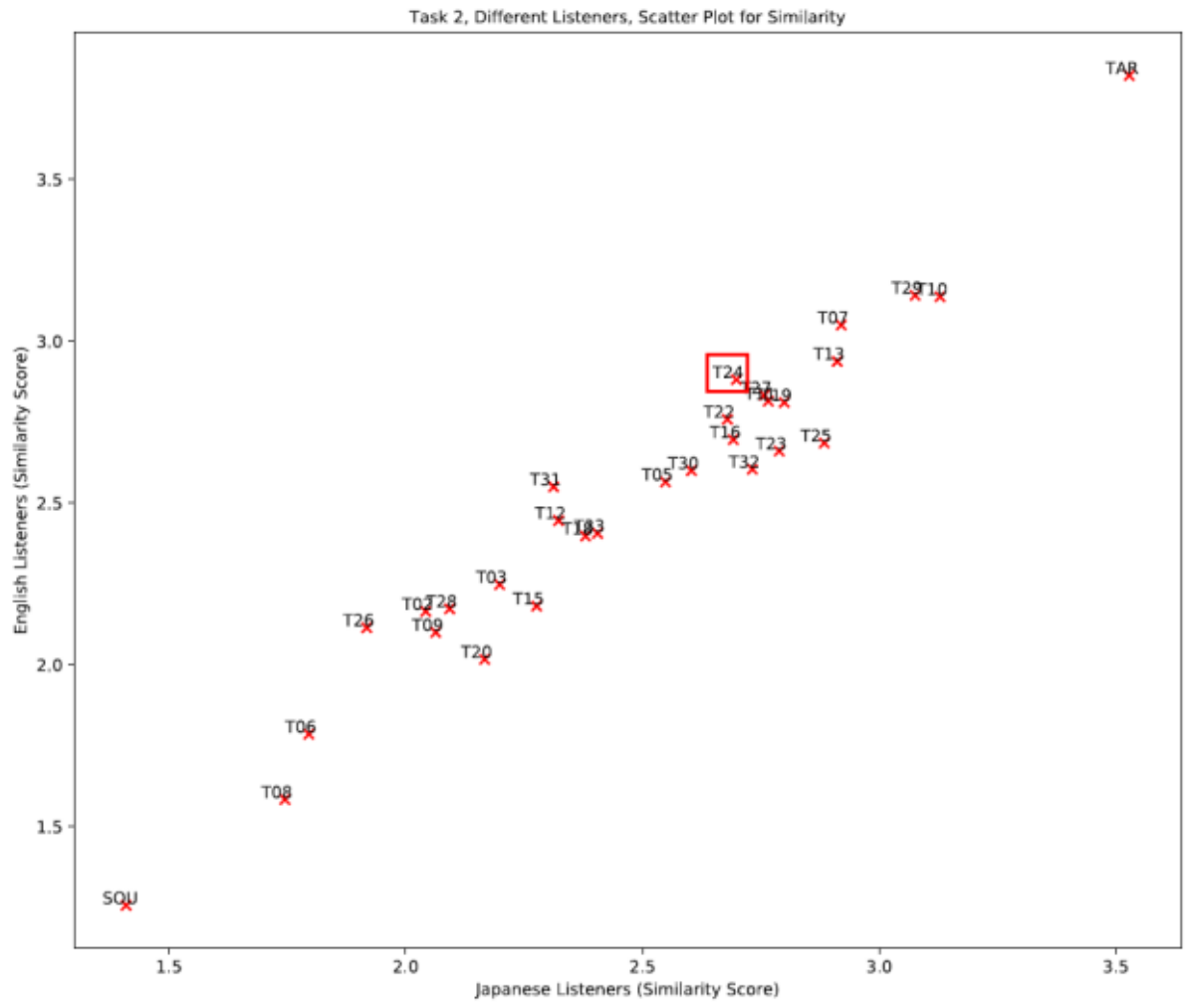

Figure 7: *Similarity MOS evaluation of task 2, from both Japanese and English listeners.*

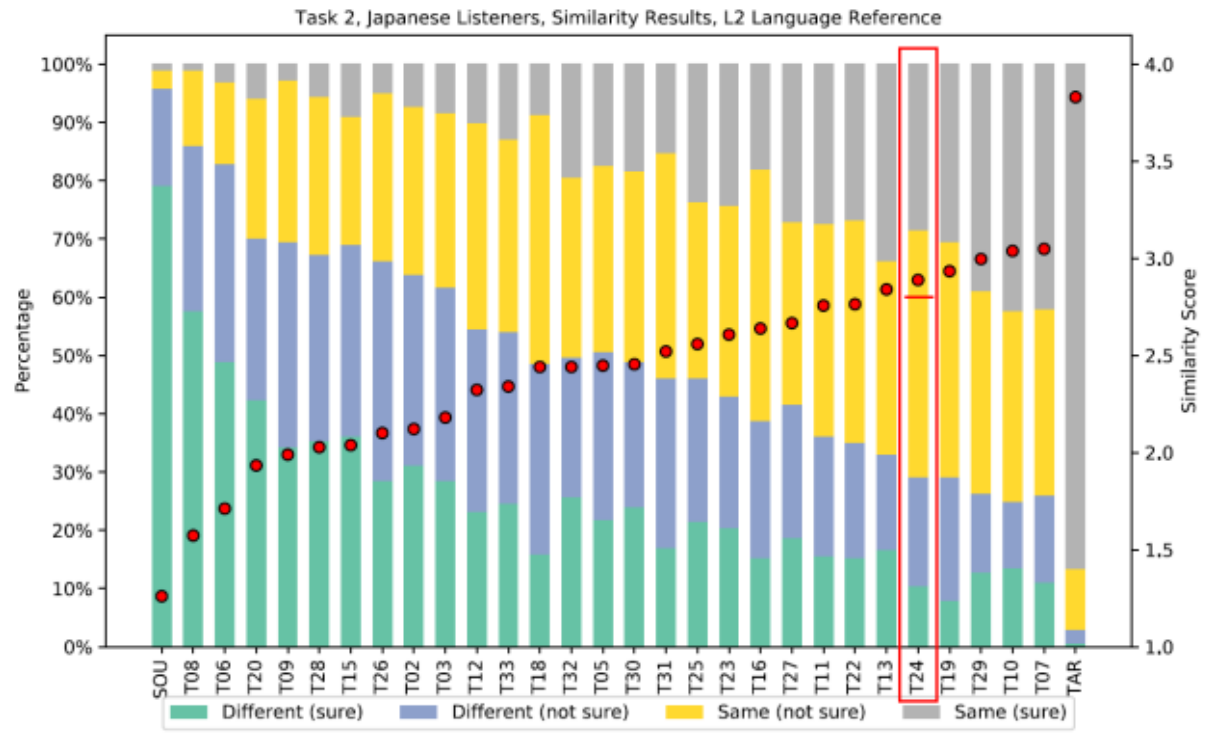

Figure 8: *Similarity results of the target speaker for task 2 when using L2 language audio as reference. Similarity scores are arranged in accordance with their mean (red dot).*

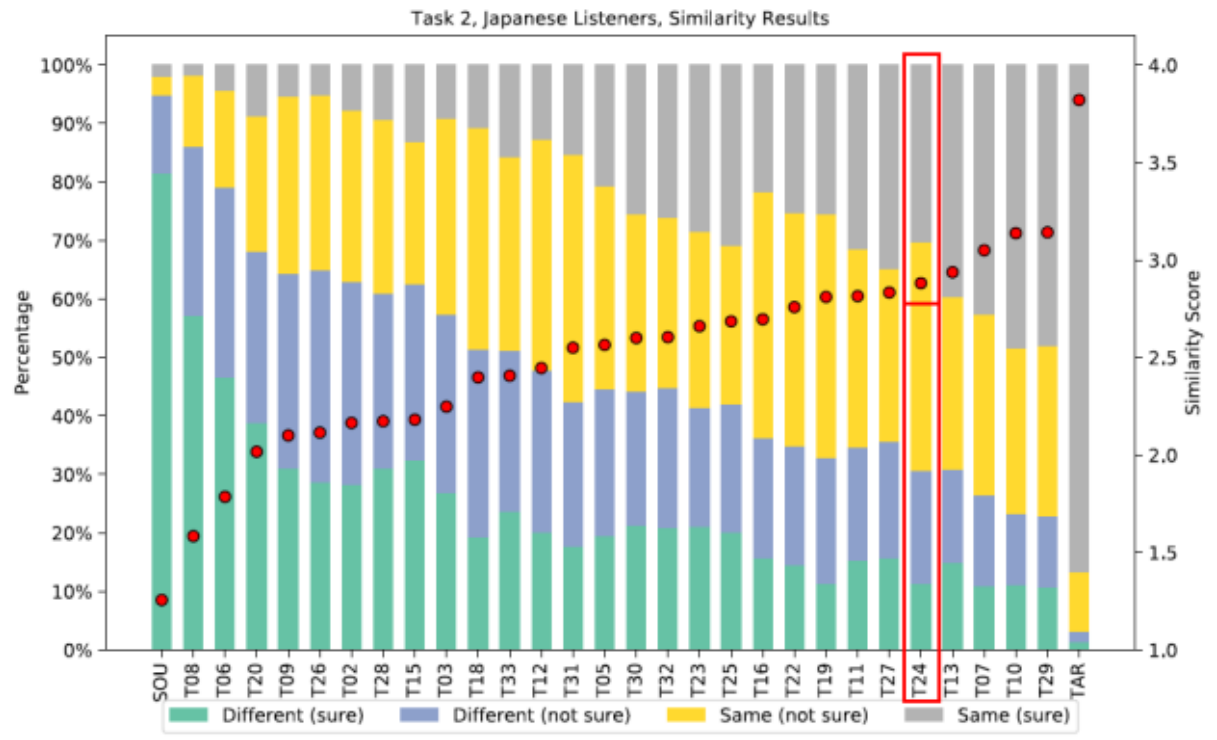

Figure 9: *Similarity results of the target speaker for task 2. Similarity scores are arranged in accordance with their mean (red dot).*

Figure 8 and Figure 9 show the similarity results of the target speaker for task 2 when using L2 language audio as reference. Similarity scores are arranged in accordance with their mean (red dot). Our result is 5th, which demonstrates the effectiveness of the Bottleneck feature in cross-lingual voice conversion and the effect of prosody in increasing the similarity.

### 4.2. Objective evalution

In this competition, the organizers added objective evaluation. This makes this VCC2020 have more indicators that can be compared. The specific indicators are shown in Table 2. As shown in Figure 10, we found that our system achieved relatively good results in ASV-based speaker similarity assessment in Task 2, ranking No. 2. This trend is consistent with the previous main observations, that is, we have a relatively good performance in Task 2 and similarity. Because it is an objective evaluation, this may reduce the sampling rate of 16k, not the negative impact of the factor of 24k. So it ranks higher than subjective evaluation. This also proves the effectiveness of our model. If we use 24k audio, we will have better results.

Table 2: *Objective evalution*

| Category | Measurement Tool | Type of Measure | Metric |
|---|---|---|---|
| **Similarity** | Automatic Speaker Verification | Target-Spoof Assessment | Equal Error Rate |
| | Automatic Speaker Verification | Converted Source and Target Similarity | Pfa (tar) |
| | Objective Mean Opinion Score | Converted Source and Source Similarity | Pmiss (src) |

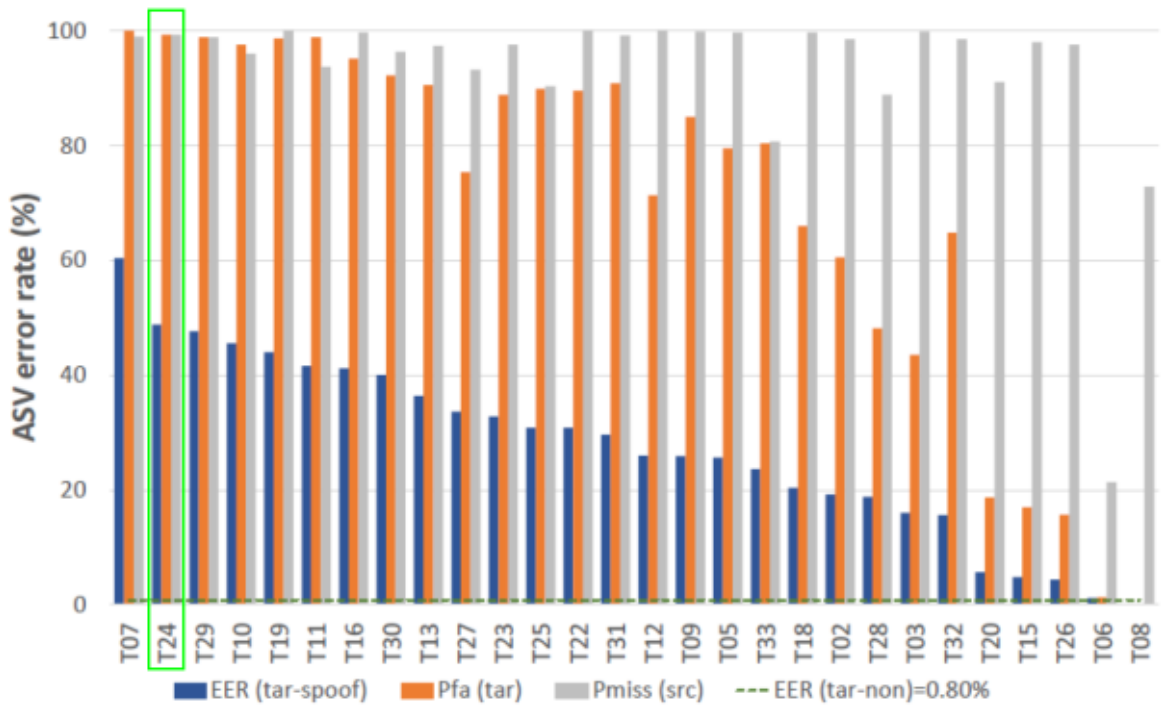

Figure 10: *ASV-based speaker similarity assessment in Task 2*

## 5. Conclusions and future work

The PPG-based end-to-end voice conversion system can now achieve a good result, and it performs quite well in terms of naturalness, similarity and best accuracy. In actual commercial use, it can be better promoted and has higher stability. It is expected that this will be one of the mainstream solutions for R. K. Das, T. Kinnunen, W.-C. Huang, Z. Ling, J. Yamagishi, Y. Zhao, X. Tian and T. Toda, "Predictions of Subjective Ratings and Spoofing Assessments of Voice Conversion Challenge 2020 Submissions," in ISCA Joint Workshop for the Blizzard Challenge and Voice Conversion Challenge 2020.voice conversion in the future, which will attract more and more researchers to study.

In the future, we will focus our work on: 1. The promotion and stability of actual commercial scenarios; 2. The improvement of pronunciation accuracy; 3. The training of small corpus or many-to-many speech conversion.

Thanks to the staff of the VCC2020 Organizing Committee! Thanks again!